\documentstyle[12pt]{article}
\begin{document}
\begin{flushright}
INFNNA-IV-2000/4
\end{flushright}
\vspace*{1.cm}

\begin{center}{RADIATIVE RARE KAON DECAYS\footnote{\small Submitted to 3rd Workshop 
on Physics and Detectors for DAPHNE (DAPHNE 99), Frascati, Italy, 16-19 Nov 1999. }}
\\
\vspace*{1.6cm}
{Giancarlo D'Ambrosio \\
\vspace*{0.4cm}
{\em INFN-Sezione di Napoli, 80126 Napoli Italy}}
\vspace*{1.6cm}
\begin{abstract}
We review some recent theoretical results on radiative rare kaon decays.
Particular attention is devoted to the channels $K\rightarrow \pi l\overline{%
l}$ and $K\rightarrow \pi \pi \gamma$, where we are (or might be) able to
extract the short distance contributions. This is achieved by a careful
study of the long distance part. We study also CP violating observables,
which are sensitive also to extensions of the SM. As byproduct, we discuss
interesting chiral tests.
\end{abstract}
\end{center}

\section{Introduction}

Rare Kaon decays are a very interesting place to test the Standard Model
(SM) and its extensions \cite{reviews,DI98}. $K\rightarrow \pi \nu \overline{
\nu }$ decays have the advantage not to be affected by long distance
uncertainties and thus they are definetely very appealing \cite{reviews,DI98}
.\ Here we study the complementary channels $K\rightarrow \pi \ell ^{+}\ell
^{-}$ and $K\rightarrow \pi \pi \gamma $ that can be studied either as
byproduct of the previous ones or also as an independent search. The long
distance contributions are in general not negligible and must be carefully
studied in order to pin down the short distance part. The advantage is that
these channels are more accessible experimentally.

\section{$K_{L}\rightarrow \pi ^{0}\ell ^{+}\ell ^{-},$ $K_{S}\rightarrow
\pi ^{0}\ell ^{+}\ell ^{-}$ and $K^{+}\rightarrow \pi ^{+}\ell ^{+}\ell ^{-}$%
}

The effective current$\otimes $current structure of weak interactions
constrains short distance contribution to $K_{L}\rightarrow \pi ^{0}\ell
^{+}\ell ^{-},$ analogously to $K_{L}\rightarrow \pi ^{0}\nu \overline{\nu }
, $ to be direct CP\ violating \cite{gino98a}. However differentely from the
neutrino case $K_{L}\rightarrow \pi ^{0}\ell ^{+}\ell ^{-}$ receives also
non-negligible long distance contributions: i) indirect $CP$ violating from 1%
$\gamma -$ exchange and ii$)$ $CP$ conserving from 2$\gamma $- exchange.
Furthermore we must warn about the danger of the potentially large
background contribution from $K_{L}\rightarrow e^{+}e^{-}\gamma \gamma $ to $%
K_{L}\rightarrow \pi ^{0}e^{+}e^{-}$ \cite{greenlee}. The present bounds
from KTeV \cite{corcoran} are 
\begin{equation}
B(K_{L}\rightarrow \pi ^{0}e^{+}e^{-})<5.6\times 10^{-10}\quad
B(K_{L}\rightarrow \pi ^{0}\mu ^{+}\mu ^{-})<3.4\times 10^{-10}.
\end{equation}

\subsection{Direct CP violating contributions}

Box+Z-penguin top loop induce the direct CP violating $K_{L}\rightarrow \pi
^{0}\ell ^{+}\ell ^{-}$ amplitude. QCD corrections have been evaluated at
next-to-leading order \cite{BL94} leading to the prediction

\[
B(K_{L}\rightarrow \pi ^{0}e^{+}e^{-})_{CPV-dir}^{SM}\ =0.69\times
10^{-10}\left[ \frac{\bar{m}_{t}(m_{t})}{170GeV}\right] ^{2}\left[ \frac{\Im
m(\lambda _{t})}{\lambda ^{5}}\right] ^{2}; 
\]
where $\lambda _{q}=\,V_{qs}^{*}V_{qd}$ and using the present constrains on $%
\Im m(\lambda _{t})$ one obtains the range \cite{reviews,BL94} 
\[
2.8\times 10^{-12}\leq B(K_{L}\rightarrow \pi
^{0}e^{+}e^{-})_{CPV-dir}^{SM}\leq 6.5\times 10^{-12}. 
\]

Lately it has been pointed out the possibility of new physics to
substantially enhance the SM\ predictions through effects that could be
parametrized by an effective dimension-4 operator $\overline{s}dZ$ vertex $%
Z_{ds}$ \cite{gino98b}$;$ the CP violating contribution $\Im m(Z_{ds})$ and
consequentely $K_{L}\rightarrow \pi ^{0}\nu \overline{\nu }$ is constrained
by the value of $\varepsilon ^{\prime }/\varepsilon $, while $\Re e(Z_{ds})$
and $K^{^{\pm }}\rightarrow \pi ^{\pm }\nu \overline{\nu }$ are limited by $%
K_{L}\rightarrow \mu \overline{\mu }$ \cite{DIP,BS99}$.\;$Also the recent
large value of $\varepsilon ^{\prime }/\varepsilon $ \cite
{KTeVeps99,NA48eps99} allows new sources of CP violating contributions \cite
{masiero99,martinelli99}. In particular, a large value for the Wilson
coefficient $C_{g}^{-}$ of the dimension-5 operators of the $|\Delta S|=1$
effective hamiltonian 
\begin{equation}
{\cal H}_{eff}^{|\Delta S|=1;d=5}=\left[ C_{\gamma }^{+}Q_{\gamma
}^{+}+C_{\gamma }^{-}Q_{\gamma }^{-}+C_{g}^{+}Q_{g}^{+}+C_{g}^{-}Q_{g}^{-}+ %
\mbox{\rm h.c.}\right] ,  \label{eq:d5}
\end{equation}
where 
\begin{equation}
Q_{\gamma }^{\pm }=\frac{Q_{d}e}{16\pi ^{2}}(\bar{s}_{L}\sigma _{\mu \nu
}d_{R}\pm \bar{s}_{R}\sigma _{\mu \nu }d_{L})\cdot F^{\mu \nu }
\label{eq:d5f}
\end{equation}
and 
\begin{equation}
Q_{g}^{\pm }=\frac{g}{16\pi ^{2}}(\bar{s}_{L}\sigma _{\mu \nu
}t^{a}G_{a}^{\mu \nu }d_{R}\pm \bar{s}_{R}\sigma _{\mu \nu }t^{a}G_{a}^{\mu
\nu }d_{L})  \label{eq:d5G}
\end{equation}
has been advocated for the large value of $\varepsilon ^{\prime
}/\varepsilon $ \cite{MM}. Indeed the SM allows only small values
(suppressed by $s,d-$quark masses) for $C_{\gamma }^{\pm }$ and $C_{g}^{\pm
} $. One can then check the consequences for $K\rightarrow \pi \ell 
\overline{\ell }$ and in general rare kaon decays of New Physics (NP) valus
for all the Wilson coefficient in (\ref{eq:d5}). Indeed one finds that $%
B(K\rightarrow \pi \ell \overline{\ell })_{NP}$ can be as much as one order
of magnitude larger than the Standard Model prediction \cite{BCIRS}. Lately
it has been pointed out that $C_{g}^{+}$ in (\ref{eq:d5}) contributes also
to the charge asymmetry in $K^{+}\rightarrow 3\pi $ \cite{DIM}.

\subsection{Indirect CP violating contribution, $K_{S}\rightarrow \pi
^{0}e^{+}e^{-}$ and $K^{\pm }\rightarrow \pi ^{\pm }l^{+}l^{-}$}

Short distance contributions to $K\rightarrow \pi \stackrel{*}{\gamma }$ are
very small compared to long distance contributions evaluated then in Chiral
Perturbation Theory ($\chi $PT ) \cite{Weinberg1}. $K\rightarrow \pi 
\stackrel{*}{\gamma }$ ($K^{\pm }\rightarrow \pi ^{\pm }\stackrel{*}{\gamma }
$ and $K_{S}\rightarrow \pi ^{0}\stackrel{*}{\gamma }$) decays start at $%
{\cal O(}p^{4})$ in $\chi $PT with loops (dominated by the $\pi \pi -$cut$)$
and counterterm contributions \cite{EPR1}.\ Higher order contributions ($%
{\cal O(}p^{6}))$ might be large, but are not completely under control since
new (and with unknown coefficients) counterterm structures appear \cite{r2}%
.\ In Ref.~\cite{r3} we have parameterized the $K\rightarrow \pi \stackrel{*%
}{\gamma }(q)$ form factor as 
\begin{equation}
W_{i}(z)\,=\,G_{F}M_{K}^{2}\,(a_{i}\,+\,b_{i}z)\,+\,W_{i}^{\pi \pi
}(z)\;,\qquad \qquad i=\pm ,S  \label{eq:ctkpg}
\end{equation}
with $z=q^{2}/M_{K}^{2}$, and where $W_{i}^{\pi \pi }(z)$ is the loop
contribution, given by the $K\rightarrow \pi \pi \pi $ unitarity cut and
completely known up to ${\cal O(}p^{6})$. All our results in that reference
are expressed in terms of the unknown parameters $a_{i}$ and $b_{i},$
expected of ${\cal O(}1)$. At the first non-trivial order, ${\cal O(}p^{4}),$
$b_{i}=0,$ while $a_{i}$ receive counterterm contributions not determined
yet. At ${\cal O(}p^{6}),$ $b_{i}\neq 0,$ while $a_{i}$ receive new and
unknown contributions. Due to the generality of (\ref{eq:ctkpg}), we expect
that $W_{i}(z)$ is a good approximation to the complete form factor. From
the $K^{+}\rightarrow \pi ^{+}e^{+}e^{-}$ experimental width and slope, E865
obtains \cite{kpeeE865} 
\begin{equation}
a_{+}\,=-0.587\pm 0.010\qquad \,b_{+}=-0.655\pm 0044  \label{eq:lpkpll}
\end{equation}
Also the fit with (\ref{eq:ctkpg}), i.e. with the genuine chiral
contribution $W_{+}^{\pi \pi }(z),$ is better ($\chi ^{2}$ /$d.o.f.$ $\sim
13.3/18)$ than just a linear slope ( $\chi ^{2}$/$d.o.f.$ $\sim 22.9/18),$
showing the validity of the chiral expansion.Then the universality of the
form factor in (\ref{eq:ctkpg}) is further tested by using (\ref{eq:lpkpll})
to predict the branching $B(K^{+}\rightarrow \pi ^{+}\mu ^{+}\mu ^{-}),\;$
which indeed perfectely agrees with the new experimental value by E865 $%
B(K^{+}\rightarrow \pi ^{+}\mu ^{+}\mu ^{-})_{{\rm exp}}=(9.22\pm 0.60\pm
0.49)\cdot 10^{-8}$\cite{kpmumuE865,lowe}. This value is however larger \cite
{r3} by $3.3$ $\sigma $ 's than the old experimental result \cite{r4}. Also
the slope in the muon channel, though with large statistical errors is now
consistent with (\ref{eq:lpkpll}) \cite{kpmumuE865}.

We should stress that it is not clear at the moment the meaning of the
apparent slow convergence in the chiral expansion in $K^{+}\rightarrow \pi
^{+}l^{+}l^{-},$ indeed the values in (\ref{eq:lpkpll}) do not respect the
naive chiral dimensional analysis expectation $\,b_{+}/a_{+}\sim
M_{K}^{2}/M_{V}^{2}.\qquad $

There is no model independent relation among $a_{S}$ and $a_{+}$ and thus a
secure determination of $B(K_{L}\rightarrow \pi
^{0}e^{+}e^{-})_{CP-indirect} $ requires a direct measurement of $%
B(K_{S}\rightarrow \pi ^{0}e^{+}e^{-}),$ possibly to be performed by KLOE at
DA$\Phi $NE \cite{r3}. The dependence from $b_{S}$ is very mild and thus we
predict $B(K_{S}\rightarrow \pi ^{0}e^{+}e^{-})\,\simeq
\,5.2\,a_{S}^{2}\,\times 10^{-9}~$. If we include the interference term
among direct and indirect the $CP$ --violating terms we obtain \cite{r3} 
\begin{equation}
B(K_{L}\rightarrow \pi ^{0}e^{+}e^{-})_{CPV}\,=\,\left[
15.3\,a_{S}^{2}\,-\,6.8\frac{\displaystyle \Im m\lambda _{t}}{\displaystyle %
10^{-4}}\,a_{S}\,+\,2.8\left( \frac{\displaystyle \Im m\lambda _{t}}{%
\displaystyle 10^{-4}}\right) ^{2}\right] \times 10^{-12}~.
\label{eq:cpvtot}
\end{equation}
A very interesting scenario emerges for $a_{S}\stackrel{<}{_{\sim }}-0.5$ or 
$a_{S}\stackrel{>}{_{\sim }}1.0$. Since $\Im m\lambda _{t}$ is expected to
be $\sim 10^{-4}$, one would have $B(K_{L}\rightarrow \pi
^{0}e^{+}e^{-})_{CPV}\stackrel{>}{_{\sim }}10^{-11}$ in this case. Moreover,
the $K_{S}\rightarrow \pi ^{0}e^{+}e^{-}$ branching ratio would be large
enough to allow a direct determination of $|a_{S}|$. Thus, from the
interference term in (\ref{eq:cpvtot}) one could perform an independent
measurement of $\Im m\lambda _{t}$, with a precision increasing with the
value of $|a_{S}|$.

\subsection{CP conserving contributions: ``$\gamma \gamma "$
inter\-media\-te sta\-te contri\-butions}

\hspace*{0.1cm} The general amplitude for $K_{L}(p)\rightarrow \pi
^{0}\gamma (q_{1})\gamma (q_{2})$ can be written in terms of two independent
Lorentz and gauge invariant amplitudes $A(z,y)$ and $B(z,y):$

\begin{eqnarray}
M^{\mu \nu } &=&\frac{\displaystyle A(z,y)}{\displaystyle m_{K}^{2}}%
\,(q_{2}^{\mu }\,q_{1}^{\nu }\,-\,q_{1}\cdot q_{2}g^{\mu \nu })\;+\;\frac{%
\displaystyle 2\,B(z,y)}{\displaystyle m_{K}^{4}}\,(-p\cdot q_{1}\,p\cdot
q_{2}\,g^{\mu \nu }\,-\,  \nonumber \\
&&\qquad \qquad -\,q_{1}\cdot q_{2}\,p^{\mu }p^{\nu }+\,p\cdot
q_{1}\,q_{2}^{\mu }p^{\nu }\,+\,p\cdot q_{2}\,p^{\mu }q_{1}^{\nu }\,\,)
\label{kpgg}
\end{eqnarray}
where $y=p\cdot (q_{1}-q_{2})/m_{K}^{2}$ and $z\,=%
\,(q_{1}+q_{2})^{2}/m_{K}^{2}$. Then the double differential rate is given
by 
\begin{equation}
\frac{\displaystyle \partial ^{2}\Gamma }{\displaystyle \partial y\,\partial
z}=\,\frac{\displaystyle m_{K}}{\displaystyle 2^{9}\pi ^{3}}[%
\,z^{2}\,|\,A\,+\,B\,|^{2}\,+\,\left( y^{2}-\frac{\displaystyle \lambda
(1,r_{\pi }^{2},z)}{\displaystyle 4}\right) ^{2}\,|\,B\,|^{2}\,]~,
\label{eq:doudif}
\end{equation}
where $\lambda (a,b,c)$ is the usual kinematical function and $r_{\pi
}=m_{\pi }/m_{K}$. Thus in the region of small $z$ (collinear photons) the $B
$ amplitude is dominant and can be determined separately from the $A$
amplitude. This feature is crucial in order to disentangle the CP-conserving
contribution $K_{L}\rightarrow \pi ^{0}e^{+}e^{-}.$ The $\gamma \gamma $
intermediate state can be i) real or ii) virtual, generating respectively an
absorptive (two-photon discontinuity) and dispersive contribution to $%
K_{L}\rightarrow \pi ^{0}e^{+}e^{-}.\;$It has been shown in a model that i)
is dominant \cite{r2}, and further support might come from the experimental%
\cite{kpeeg} and theoretical \cite{Gabbiani97} study of $K_{L}\rightarrow
\pi ^{0}e^{+}e^{-}\gamma .$

The two photons in the $A$-type amplitude are in a state of total angular
momentum $J=0$ ($J,$ total diphoton angular momentum), and it turns out that
for this contribution $A(K_{L}\rightarrow \pi ^{0}e^{+}e^{-})_{J=0}\sim
m_{e} $ ($m_{e}$ electron mass) \cite{EPR88}; however the higher angular
momentum state $B$-type amplitude in (\ref{kpgg}), though chirally and
kinematically suppressed for $A(K_{L}\rightarrow \pi ^{0}\gamma \gamma ),$
generate $A(K_{L}\rightarrow \pi ^{0}e^{+}e^{-})_{J\neq 0}$ competitive with
the CP violating contributions \cite{r2}.

The leading finite ${\cal O}(p^{4})$ amplitudes of $K_{L}\rightarrow
\pi^{0}\gamma \gamma $ generates only the $A$--type amplitude in Eq.~(\ref
{eq:doudif}) \cite{Cappiello}. This underestimates the observed branching
ratio, $(1.68\pm 0.07\pm 0.08)\times 10^{-6}$ \cite{KTeVkpgg99} by a large
factor but reproduces the experimental spectrum, predicting no events at
small $z$. The two presumably large ${\cal O}(p^{6})$ contributions have
been studied: i) the ${\cal O}(p^{6})$ unitarity corrections \cite
{CD93,CE93,KH94} that enhance the ${\cal O}(p^{4})$ branching ratio by $40\%$
and generate a $B$ --type amplitude, ii) the vector meson exchange
contributions that are in general model dependent \cite{SE88,EP90} but can
be parameterized $K_{L}\rightarrow \pi ^{0}\gamma \gamma $ by an effective
vector coupling $a_{V}$ \cite{EP90} . Then the contribution to $%
K_{L}\rightarrow \pi ^{0}e^{+}e^{-}$ is determined by the value of $a_{V}.$
The agreement with experimental $K_{L}\rightarrow \pi ^{0}\gamma \gamma $
rate and spectrum would demand $a_{V}\sim -0.8\;$\cite{CE93,DP97}.

It would be desirable to have a theoretical understanding of this value.\
Indeed we have related $a_{V}$ with the $K_{L}\rightarrow \gamma \gamma ^{*}$
linear slope, $b$ \cite{DP97}; the experimental value is $%
b_{exp}\,=\,0.81\pm 0.18\,$. Theoretically the slope $b$ is also generated
by vector meson exchange contribution.

We can evaluate now $a_{V}$ and the $K_{L}\rightarrow \gamma \gamma ^{*}$
slope $b$ in factorization (FM), i.e. writing a {\it current}$\times ${\it %
current }structure 
\begin{equation}
{\cal L}_{FM}\;=\;4\,k_{F}\,G_{8}\,\langle \,\lambda \,J_{\mu }\,J^{\mu
}\,\rangle \;+\;h.c.\;\;\;,  \label{FM}
\end{equation}
where $\lambda \,\equiv \,\frac{1}{2}\,(\lambda _{6}\,-\,i\lambda _{7}),$ $%
G_{8}$ is determined from $K\rightarrow \pi \pi $ and the fudge factor $%
k_{F}\sim {\cal O}(1)$ has to be determined phenomenogically. A satisfactory
understanding of the model would require $k_{F}\sim 0.2-0.3,$ to match the
perturbative result.

There are two ways to derive the FM weak lagrangian generated by resonance
exchange (this corresponds to different ways to determine the conserved
current $J_{\mu }$) \cite{DP97,DP98}~:

\begin{itemize}
\item[(${\cal A}$)]  To evaluate the strong action generated by resonance
exchange, and then perform the factorization procedure in Eq.~(\ref{FM}). By
this way, since we apply the FM procedure once the vectors have already been
integrated out the lagrangian is generated at the kaon mass scale.

\item[(${\cal B}$)]  Otherwise, we can first write down the spin--1 strong
and weak chiral lagrangian.\ The weak resonance coupling constants are
determined in factorization. We then integrate out the resonance fields so
that the effective lagrangian is generated at the scale of the resonance.
\end{itemize}

In principle the two effective actions do not coincide and phenomenology may
prefer one pattern \cite{DP98}.\ In the case at hand, ${\cal A}$ and ${\cal B%
}$ give different structures, however they both generate a good
phenomenology with one free\ parameter $k_{F}$ , i.e.

\begin{equation}
a_{V}\,\simeq \,-0.72\;\;\;\;\;,\;\;\;\;\;b\,\simeq \,0.8-0.9~,
\end{equation}
but with different value of $k_{F}:$ ${\cal A}$ $\Rightarrow $ $k_{F}=1$
while ${\cal B}$ $\Rightarrow $ $k_{F}=0.2.$ Interestingly this seems to
suggest that the matching should be performed at the resonance scale.

Very interestingly the new data from KTeV \cite{KTeVkpgg99} confirms sharply
our prediction for $a_{V}:$ $a_{V}=-0.72\pm 0.05\pm 0.06$ and show a clear
evidence of events at low z. This turns in a stringent limit for the CP
conserving contribution to $K_{L}\rightarrow \pi ^{0}e^{+}e^{-}$: $%
1.<B(K_{L}\rightarrow \pi ^{0}e^{+}e^{-})\cdot 10^{12}<4$ \cite{r2,DP97}.
The direct measurement of the events at low z will give a direct, model
independent and precise determination of the CP conserving contribution to $%
K_{L}\rightarrow \pi ^{0}e^{+}e^{-}.$

\section{$K\to \pi \pi \gamma $}

The $K(p)\rightarrow \pi (p_{1})\pi (p_{2})\gamma (q)$ amplitude is usually
decomposed also in electric ($E)$ and the magnetic ($M)$ terms \cite{DI98}.
Defining the dimensionless amplitudes $E$ and $M$ as in \cite
{DMS93,ecker94,DI98}, we can write: 
\begin{equation}
A(K\to \pi \pi \gamma )=\varepsilon _{\mu }(q)\left[ E(z_{i})(p_{1}\cdot
q\,p_{2}^{\mu }-p_{2}\cdot q\,p_{1}^{\mu })+M(z_{i})\epsilon ^{\mu \nu \rho
\sigma }p_{1\nu }p_{2\rho }q_{\sigma }\right] /m_{K}^{3},  \label{amplitude}
\end{equation}
where 
\begin{equation}
z_{i}={\frac{p_{i}\cdot q}{m_{K}^{2}}},\qquad \qquad {\rm and}\qquad \qquad
z_{3}={\frac{p_{_{K}}\cdot q}{m_{K}^{2}}}=z_{1}+z_{2}.
\end{equation}
In the electric transitions one generally separates the bremsstrahlung
amplitude $E_{B}$ : if $eQ_{i}$ is the electric charge of the pion $\pi $%
\begin{equation}
E_{B}(z_{i})\doteq {\frac{eA(K\to \pi _{1}\pi _{2})}{M_{K}z_{3}}}\left( {%
\frac{Q_{2}}{z_{2}}}-{\frac{Q_{1}}{z_{1}}}\right) .  \label{bremss1}
\end{equation}
$E_{B}$ is generally enhanced due to the factor $1/E_{\gamma }^{*}$ for $%
E_{\gamma }^{*}\rightarrow 0,$ where $E_{\gamma }^{*} $ is the photon energy
in the kaon rest frame. Summing over photon helicities there is no
interference among electric and magnetic terms: 
\begin{eqnarray}
{\frac{\mbox{d}\Gamma }{\mbox{d}z_{1}\mbox{d}z_{2}}} &=&{\frac{M_{K}}{4(4\pi
)^{3}}}\left( |E(z_{i})|^{2}+|M(z_{i})|^{2}\right)  \nonumber \\
&&\times \left[
z_{1}z_{2}(1-2z_{3}-r_{1}^{2}-r_{2}^{2})-r_{1}^{2}z_{2}^{2}-r_{2}^{2}z_{1}^{2}\right] ,
\label{kppgw}
\end{eqnarray}
($r_{m}=m_{\pi }/m_{K}$). At the lowest order ($p^{2})$ in $\chi PT$ one
obtains only $E_{B}$

Magnetic and electric direct emission amplitudes, appearing at ${\cal O(}%
p^{4}),$ can be decomposed in a multipole expansion (see Ref.\cite
{LV88,DMS93,DI98}). In the table below we show the present $K\to \pi \pi
\gamma $ experimental status; also shown are the reason for the suppression
of the bremsstrahlung amplitude and the leading multipole amplitudes.

\noindent $
\begin{array}[t]{|c|c|c|}
\hline
\mbox{decay} & BR(\mbox{bremsstrahlung}) & BR(\mbox{direct 
emission}) \\ \hline
\begin{array}{l}
K^{\pm }\to \pi ^{\pm }\pi ^{0}\gamma \\ 
T_{\pi ^{+}}^{*}=(55-90)MeV
\end{array}
& 
\begin{array}{c}
(2.57\pm 0.16)\times 10^{-4} \\ 
(\Delta I=3/2)
\end{array}
& 
\begin{array}{c}
(4.72\pm 0.77)\times 10^{-6} \\ 
E1,M1
\end{array}
\\ \hline
\begin{array}{l}
K_{S}\to \pi ^{+}\pi ^{-}\gamma \\ 
E_{\gamma }^{*}>50MeV
\end{array}
& (1.78\pm 0.05)\times 10^{-3} & <6\times 10^{-5}(E1) \\ \hline
\begin{array}{l}
K_{L}\to \pi ^{+}\pi ^{-}\gamma \\ 
E_{\gamma }^{*}>20MeV
\end{array}
\quad & 
\begin{array}{c}
(1.49\pm 0.08)\times 10^{-5} \\ 
(CP{\rm \ violation)}
\end{array}
& 
\begin{array}{c}
(3.09\pm 0.06)\times 10^{-5} \\ 
M1,E2
\end{array}
\\ \hline
K_{S}\to \pi ^{0}\pi ^{0}\gamma &  & M2 \\ \hline
K_{L}\to \pi ^{0}\pi ^{0}\gamma & <5.6\times 10^{-6} & E2 \\ \hline
\end{array}
$

\medskip

We do not discuss $K_{S,L}\to \pi ^{0}\pi ^{0}\gamma $ due to the small
branching ratio ($<10^{-8})$\cite{HS93}. $K_{S}\to \pi ^{+}\pi ^{-}\gamma $
has been discussed in \cite{DMS} and no new experimental results have been
reported recentely. While motivated by new results we update $%
K_{L}\rightarrow \pi ^{+}\pi ^{-}\gamma $ and $K^{+}\rightarrow \pi ^{+}\pi
^{0}\gamma .$

\underline{$K_{L}\rightarrow \pi ^{+}\pi ^{-}\gamma $($\gamma ^{*})$}: The
bremsstrahlung ($E_{B})$ is suppressed by $CP$ violation\ ($\sim \eta _{+-}$
) and firmly established theoretically from (\ref{bremss1}) predicting $%
B(K_{L}\rightarrow \pi ^{+}\pi ^{-}\gamma )_{IB}=1.42\cdot 10^{-5}$ \cite
{Eckerrep}. This contribution has been also measured by interference with
the $M1$ transition in $K_{L}\rightarrow \pi ^{+}\pi ^{-}e^{+}e^{-}$ \ \cite
{belz99,Savage}. Due to the large slope, KTeV parametrizes the magnetic
amplitude in (\ref{amplitude}) as $e{\cal F}/M_{K}^{4}$ and

\begin{equation}
{\cal F=}\widetilde{g}_{M1}\left[ \frac{a_{1}}{(M_{\rho
}^{2}-M_{K}^{2})+2M_{K}E_{\gamma }^{*}}+a_{2}\right]
\label{amplitudeKLpipig}
\end{equation}
finding $a_{1}/a_{2}=(-0.729\pm 0.026($stat))$GeV^{2}$ and the branching
given in the table, which fixes also $\widetilde{g}_{M1}$.\ Such large slope
can be accomodated in various Vector dominance schemes \cite
{HS93,ecker94,kppgVMD}, while the rate is very sensitive to $SU(3)-$breaking
and unknown $p^{4}$ unknown low energy contributions and thus difficult to
predict.

\underline{$K^{+}\rightarrow \pi ^{+}\pi ^{0}\gamma $}: New data from BNL\
E787\cite{Komatsubara99} show van\-ish\-ing in\-ter\-fer\-ence among
bremss\-trahlung and elec\-tric tran\-si\-tion. Thus the direct emission
branching ($B(K^{+}\rightarrow \pi ^{+}\pi ^{0}\gamma $)$_{{\rm exp}}^{DE}),$
in the table, must be interpreted as a pure magnetic transition.
Theoretically one can identify two different sources for $M,$ appearing at $%
{\cal O(}p^{4}):$ i) a pole diagram with a Wess-Zumino term and ii) a pure
weak contact term, generated also in factorization by an anomalous current 
\cite{CH90a,BE92}. \ $B(K^{+}\rightarrow \pi ^{+}\pi ^{0}\gamma $)$_{{\rm exp%
}}^{DE}$ is substantially smaller than previous values, but still show that
the contribution ii) is non-vanishing.

\subsection{CP Violation}

Direct $CP$ violation can be established in the width charge asymmetry in $%
K^{\pm }\rightarrow \pi ^{\pm }\pi ^{0}\gamma ,$ $\delta \Gamma /2\Gamma $
and \ in the interference $E_{B}$ with $E_{1}$ in $K_{L}\rightarrow \pi
^{+}\pi ^{-}\gamma $ ($E$ with $M_{1}$ in $K_{L}\rightarrow \pi ^{+}\pi
^{-}e^{+}e^{-})$; both observables are kinematically difficult since one is
looking for large photon energy distribution\cite{DI98}. SM charge asymmetry
were looked in \cite{CPold} expecting $\delta \Gamma /2\Gamma \leq 10^{-5}$.
General bounds on new physics in $M1$ transitions have been studied in \cite
{HV99}, while the effects of dimension-5 operators in (\ref{eq:d5}) $E1$
transitions have been studied in \cite{CIP}, where for instance it has been
shown that the value of $\Re \left( \varepsilon ^{\prime }/\varepsilon
\right) $ allows in particular kinematical regions a factor 10\ larger than
SM.

\section{Conclusions}

We think that the recent experimental results in $K$ decays, for instance $%
\varepsilon ^{\prime }/\varepsilon $ and $K^{+}\rightarrow \pi
^{+}l^{+}l^{-},$ encourage us to think that stringent tests of the SM and of
its possible extensions are not too far ahead. From the theoretical side,
radiative rare kaon decays can be a good laboratotory to understand very
interesting questions like why the size of $b_{+}/a_{+}$ in (\ref{eq:lpkpll}%
) does not respect chiral counting. Interestingly a similar question for $%
K_{L}\rightarrow \pi ^{0}\gamma \gamma $ got finally an answer, as we have
shown in Section 2.3: the full $K\rightarrow 3\pi$ amplitude and VMD enhance
the ${\cal O(}p^{6})$ contributions.\ May be we have just to work harder for 
$K^{+}\rightarrow \pi ^{+}l^{+}l^{-}$, but at the end we may get predictive
power and also interesting physics insight. For the people who think that
theorists find always a good excuse I remind $K_{S}\rightarrow \gamma \gamma 
$ \cite{DEG}$,$ where theory is very predictive and no large higher order
contributions can be found. Finally interesting analytic approaches to weak
matrix elements has been recently exploited in Ref.\cite{derafael}, where
the relevant Green functions are evaluated and matched.

\section{Acknowledgements}

I am happy to thank the organizers for the nice atmosphere of the Workshop.
I also enjoyed working and discussing with G. Buchalla, G. Ecker, G.
Isidori, H. Neufeld and J. Portoles.\ This work is sup\-ported in part by
TMR, EC--Contract No. ERBFMRX--CT980169 (EURODA$\Phi $NE).

\end{document}